\documentstyle[12pt,epsfig,twoside]{article}
\epsfverbosetrue
\newcommand{\beq}{\begin{equation}}
\newcommand{\eeq}{\end{equation}}
\newcommand{\beqn}{\begin{eqnarray}}
\newcommand{\eeqn}{\end{eqnarray}}
\newcommand{\beqns}{\begin{eqnarray*}}
\newcommand{\eeqns}{\end{eqnarray*}}
\newcommand{\vs}{\\[0.3cm]\indent}

\newcommand{\hm}{\hspace{-0.15cm}}
\newcommand{\hsm}{\hspace{-0.2cm}}
\newcommand{\intl}{\int\limits}
\newcommand{\ointl}{\oint\limits}
\newcommand{\mc}{\multicolumn}

\def\NP{{\it Nucl. Phys.}}
\def\PL{{\it Phys. Lett.}}
\def\PR{{\it Phys. Rev.}}

\def\PRL{{\it Phys. Rev. Lett.}}
\def\NIM{{\it Nucl. Inst. Meth.}}
\def\ZP{{\it Z. Phys.}}

\def\EPJC{{\it The Europ. Phys. J. C}}
\def\JP{{\it J. Phys.}}

\def\ea{{\it et al.}}
\def\Cl{Collaboration}
\def\sf{spectral function}
\def\sfs{spectral functions}

\def\SFs{Spectral Functions}
\def\asm{$\alpha_s(M_\tau^2)$}

\def\br{branching ratio}
\def\brs{branching ratios}

\def\ie{{\it i.e.}} 
\def\eg{{\it e.g.}} 
\def\via{via} 

\begin{document}

\begin{titlepage}
\setcounter{page}{1}

\begin{flushright} 
{\bf IPNO/TH 98-05}\\
{\bf LAL 98-05}\\
February 1998
\end{flushright} 

\begin{center}
\vspace{1cm}
{\Large
  {\LARGE F}INITE {\LARGE E}NERGY {\LARGE C}HIRAL 
  {\LARGE S}UM {\LARGE R}ULES \\[0.3cm]
   AND $\tau$ {\LARGE S}PECTRAL {\LARGE F}UNCTIONS \\
}
\vspace{1.6cm}
\begin{large}
M.~Davier$^{\rm a,\,}$\footnote{E-mail: davier@lal.in2p3.fr},
L.~Girlanda$^{\rm b,\,}$\footnote{E-mail: girlanda@ipno.in2p3.fr}, 
A.~H\"ocker$^{\rm a,\,}$\footnote{E-mail: hoecker@lal.in2p3.fr} 
and J.~Stern$^{\rm b,\,}$\footnote{E-mail: stern@ipno.in2p3.fr} \\
\end{large}
\vspace{0.5cm}
{\small \em $^{\rm a}$Laboratoire de l'Acc\'el\'erateur Lin\'eaire,\\
IN2P3-CNRS et Universit\'e de Paris-Sud, F-91405 Orsay, France}\\
\vspace{0.5cm}
{\small \em $^{\rm b}$Division de Physique Th\'eorique, 
                      Institut de Physique Nucl\'eaire\\
Universit\'e de Paris-Sud, F-91406 Orsay, France}\\
\vspace{2.2cm}

{\small{\bf Abstract}}
\end{center}
{\small
\vspace{-0.2cm}
A combination  of f\/inite energy sum rule techniques and 
Chiral Perturbation Theory ($\chi$PT) is used in order to exploit 
recent ALEPH data on the non-strange $\tau$ vector $(V)$ and 
axial-vector $(A)$ \sfs\ with respect to an experimental 
determination of the $\chi$PT quantity $L_{10}$. A constrained f\/it 
of $R_{\tau,V-A}^{(k,l)}$ inverse moments $(l<0)$ and positive 
spectral moments $(l\ge0)$ adjusts simultaneously $L_{10}$ and 
the nonperturbative power terms of the Operator Product Expansion. 
We give explicit formulae for the f\/irst $k=0,1$ and $l=-1,-2$ strange 
and non-strange inverse moment chiral sum rules to one-loop order 
generalized $\chi$PT. Our f\/inal result reads 
$L^r_{10}(M_{\rho})=-(5.13\pm0.19)\times10^{-3}$, where the error includes 
experimental and theoretical uncertainties. 
\vspace{2cm}
\noindent
}
\vfill
\centerline{\it (Submitted to Physics Letters B)}
\vspace{1cm}
\thispagestyle{empty}

\end{titlepage}

\newpage\thispagestyle{empty}{\tiny.}\newpage
\setcounter{page}{1}
\setcounter{footnote}{0}
%
%
\section{Introduction}

The nonperturbative features of strong interactions make QCD a rich
environment for theoretical investigations. At suf\/f\/iciently high
energies  it is possible to parametrize the
nonperturbative ef\/fects by vacuum condensates, following the rules
of Wilson's Operator Product Expansion (OPE)~\cite{wilson}.
The universal character of these condensates has been used in the
derivation of the so-called  QCD spectral sum rules~\cite{svz}
allowing, in principle, their determination from experiment.
A particular role is played by the condensates which are order parameters of
the spontaneous breakdown of chiral symmetry (SB$\chi$S). The latter vanish
at all orders of perturbation theory and they control the high energy
behavior of chiral correlation functions, such as the dif\/ference of vector
and axial current two-point functions. On the  other hand, at low energies, 
SB$\chi$S  makes it possible to construct an ef\/fective theory
of QCD, the Chiral Perturbation Theory ($\chi$PT)~\cite{wein,gale1},
which uses  the Goldstone bosons as  fundamental f\/ields and provides a
systematic expansion of QCD correlation functions in powers of momenta and
quark masses. All missing
information is then parametrized by low-energy coupling constants,
which can be determined phenomenologically in low-energy experiments
involving  pions and kaons. The fundamental parameters describing chiral
symmetry breaking, the running quark masses and the quark anti-quark
condensates $\langle \bar q q \rangle$ appear both in low-energy ($\chi$PT)
and the high energy OPE expansion. For this reason it is useful to combine
the two expansions in order to get a truly systematic approach to the 
chiral sum rules~\cite{chiralsr}. In this paper the combined  approach 
is illustrated  through a determination of the $L_{10}$ constant of the 
chiral lagrangian, including high-energy corrections coming from the 
OPE. The connection between the two domains is provided by 
experimental data on $\tau$ hadronic \sfs\ published recently by the 
ALEPH Collaboration~\cite{aleph_vsf,aleph_asf}.
\vs
At the leading  order of $\chi$PT, $L_{10}$ is directly linked to the 
vector, $v_1$, and axial-vector, $a_1$, spin-one \sfs\ (the subscripts 
refer to the spin $J$ of the hadronic system) through the Das-Mathur-Okubo 
sum rule~\cite{dmo}  
\beq
\label{eq_dmo}
   \frac{1}{4\pi^2}\hsm\intl_0^{s_0\rightarrow\infty}\hsm
             ds\,\frac{1}{s}\left[v_1(s)-a_1(s)\right]
        \simeq -4L_{10}~.
\eeq
As it stands the DMO sum rule~(\ref{eq_dmo}) is subject to chiral
corrections due to non-vanishing quark masses~\cite{kagodmo}. On the 
other hand, the integral has to be cut at some f\/inite energy 
$s_0\leq M_{\tau}^2$, since no experimental information on $v_1-a_1$ 
is available above $M_{\tau}^2$. This truncation introduces an error 
which competes with the low-energy chiral corrections. Both types of 
corrections can be systematically included through $i)$~the high-energy 
expansion in $\alpha_s(s_0)$ and in inverse powers of $s_0$, and 
$ii)$~the low-energy expansion in powers of quark masses and of 
their logarithms.
%
%
\section{Spectral Moments}

\begin{figure}[t]
  \epsfxsize9.cm
  \centerline{\epsffile{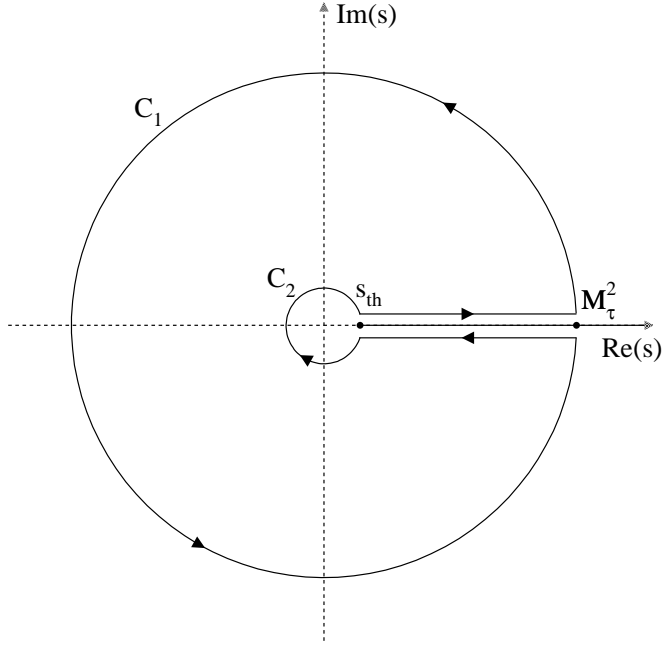}}
  \caption[.]{\label{fig_contour}\it
              Integration contour around the circles at $s=M_\tau^2$
              and $s=s_{\rm th}$.}
\end{figure}

Using unitarity and analyticity, the \sfs\ are connected 
to the imaginary part of the two-point correlation functions,
\beqn
\label{eq_correlator}
   \Pi_{ij,U}^{\mu\nu}(q) 
      &\equiv &
        i\int d^4x\,e^{iqx}
        \langle 0|T(U_{ij}^\mu(x)U_{ij}^\nu(0)^\dag)|0\rangle \nonumber \\
      &=& 
        (-g^{\mu\nu}q^2+q^\mu q^\nu)\,\Pi^{(1)}_{ij,U}(q^2)
        +q^\mu q^\nu\,\Pi^{(0)}_{ij,U}(q^2)~,
\eeqn
of vector $(U_{ij}^\mu\equiv V_{ij}^\mu=\bar{q}_j\gamma^\mu q_i)$
or axial-vector 
$(U_{ij}^\mu\equiv A_{ij}^\mu=\bar{q}_j\gamma^\mu\gamma_5 q_i)$
colour-singlet quark currents 
for time-like momentum-squared $q^2>0$. Lorentz decomposition 
is used to separate the correlation function into its $J=1$ and 
$J=0$ parts. The correlation function~(\ref{eq_correlator}) is
analytic everywhere in the complex $s$ plane except on the 
positive real axis where singularities exist. Using the 
def\/initions adopted in Refs.~\cite{aleph_vsf,aleph_asf} together 
with Eq.~(\ref{eq_correlator}), one identif\/ies for non-strange 
quark currents 
\beq
\label{eq_imv}
   {\rm Im}\,\Pi^{(1)}_{\bar{u}d,V/A}(s) = \frac{1}{2\pi}v_1/a_1(s)~,
   \hspace{1cm}
   {\rm Im}\,\Pi^{(0)}_{\bar{u}d,A}(s)   = \frac{1}{2\pi}a_0(s)~.
\eeq
Due to the conserved vector current, there is no $J=0$ contribution 
to the vector \sf, while the only contribution to $a_0$ is assumed 
to be from the pion pole. It is connected \via\ PCAC to the pion decay 
constant, $a_{0,\,{\pi}}(s)=4\pi^2 f_\pi^2\,\delta(s-m_\pi^2)$.
\vs
According to the method proposed by Le Diberder and Pich~\cite{pichledib}, 
it is possible to exploit the information from the explicit shape of the 
\sfs\ by calculating so-called spectral moments, \ie, weighted integrals 
over the \sfs. If $W(s)$ is an analytic function, by Cauchy's theorem, 
the imaginary part of $\Pi_{ij,{\rm{V/A}}}^{(J)}$ is proportional to the 
discontinuity across the positive real axis:
\beq
\label{eq_cauchy}
    \intl_0^{s_0}ds\,W(s){\rm Im}\Pi^{(J)}_{ij,{\rm{V/A}}}(s) 
       = -\frac{1}{2i}\hm\ointl_{|s|=s_0}\hm ds\,
                      W(s)\Pi^{(J)}_{ij,{\rm{V/A}}}(s)~,
\eeq
where $s_0$ is large enough for the OPE series to converge.
The authors of~\cite{pichledib} choose for $W(s)$ the functions
\beq
\label{eq_moments}
    W^{(k,l)}(s) = \left(1 - \frac{s}{s_0} \right)^{\!\!2+k} 
                \left(\frac{s}{s_0}\right)^{\!\!l}~,
\eeq
with $k$ and $l$ positive integers. The factor $(1-s/s_0)^k$ 
suppresses the integrand at the crossing of the positive real 
axis where the validity of the OPE is questioned. Its counterpart 
$(s/s_0)^l$ projects on higher energies. These moments were successfully 
applied in order to constrain nonperturbative contributions to the 
$\tau$ hadronic width, $R_\tau$, a procedure  which lead to precise 
determinations of \asm~\cite{aleph_asf,aleph_as,cleo_as}.
\vs
The extension of the spectral moment analysis to negative integer 
values of $l$ (``inverse moment sum rules'', (IMSR)~\cite{imsr})
requires, due to the pole at $s=0$, a modif\/ied contour of integration 
in the complex $s$ plane, as shown in Fig.~\ref{fig_contour}. This is 
where $\chi$PT comes into play: along the small circle placed at the 
production threshold, $s_{\rm th}=4M_\pi^2$ for non-strange $(\bar{u}d)$ 
and $s_{\rm th}=(M_\pi + M_K)^2$ for strange $(\bar{u}s)$ currents, we 
can use $\chi$PT predictions for the two-point correlators. 
Using the weight function~(\ref{eq_moments}) we adopt the following 
def\/inition of the moments:
\beqn
\label{eq_momtheory}
     \lefteqn{R_{\tau,V/A}^{(k,l)} \equiv
          12\pi|V_{ud}|^2 S_{\rm EW}\intl_{s_{\rm{min}}}^{M_\tau^2}
                        \frac{ds}{M_\tau^2}
                        \left(1 - \frac{s}{M_\tau^2}\right)^{\!\!2+k}
                        \left(\frac{s}{M_\tau^2}\right)^{\!\!l}} \nonumber \\
        & &\hspace{4cm} \times
                        \left[\left(1 + 2\frac{s}{M_\tau^2}\right)
                              {\rm Im}\Pi_{V/A}^{(0+1)}(s) 
                              - 2\frac{s}{M_\tau^2}{\rm Im}\Pi_{A}^{(0)}(s)
                        \right]~, 
\eeqn
where $s_{\rm min}=0$ for the positive moments\footnote
{
  This is due to the pion pole which is at zero mass in the chiral 
  limit.
}
$(l\ge0)$ and $s_{\rm min}=s_{\rm th}$, which is the continuum threshold, 
for the inverse moments. According to the relation~(\ref{eq_cauchy}), 
Eq.~(\ref{eq_momtheory}) reads
\beqn 
\label{eq_momtheorycontour}
   R_{\tau,V/A}^{(k,l)} 
         &=& 6 \pi i |V_{ud}|^2 S_{\rm EW}
           \ointl_C \frac{ds}{M_{\tau}^2} 
                    \left(1 - \frac{s}{M_{\tau}^2} \right)^{\!\!2+k}
                    \left(\frac{s}{M_{\tau}^2} \right)^{\!\!l} \nonumber \\
        & &\hspace{2.8cm} \times
                    \left[\left(1 + 2\frac{s}{M_{\tau}^2}
                          \right)\Pi^{(0+1)}_{V/A} (s) - 2
                          \frac{s}{M_{\tau}^2} \Pi^{(0)}_{A}(s) 
                    \right]~,
\eeqn
where $C=C_1+C_2$ for the inverse moments and $C=C_1$ for the 
positive moments (see F\/ig.~\ref{fig_contour}).
\vs
Due to the cut of the integral~(\ref{eq_momtheory}) at $M_\tau^2$,
nonperturbative physics parametrized by the short-distance OPE for 
scalar operators~\cite{wilson,svz,bnp} must be considered:
\beq
\label{eq_ope}
    \Pi_{V/A}^{(J)}(s) = 
        \hm\sum_{D=0,2,4,\dots}\frac{1}{(-s)^{D/2}}
        \hm\sum_{{\rm dim}{\cal O}=D}\hm C_{V/A}^{(J)}(s,\mu^2)
                      \langle{\cal O}_{V/A}(\mu^2)\rangle~.
\eeq
The parameter $\mu$ separates the long-distance nonperturbative 
ef\/fects, absorbed into the vacuum expectation elements 
$\langle{\cal O}_{V/A}(\mu^2)\rangle$, from the short-distance 
ef\/fects which are included in the Wilson coef\/f\/icients 
$C_{V/A}(s,\mu^2)$~\cite{wilson}. We will assume the convergence 
of the OPE series at the $\tau$ mass. This is justif\/ied in the 
light of the success of the analysis performed in 
Ref.~\cite{aleph_asf} (see Ref.~\cite{thesis} for details). 
Using the formulae of Refs.~\cite{bnp} and~\cite{chetyrkin} for 
the nonperturbative power expansion of the correlators, one 
obtains for the $(V-A)$ case
\beqn
\label{eq_corr10}
     \Pi_{\bar u d, V-A}^{(0+1)}(-s) 
       &=& 
             -\, \frac{a_s(s)}{\pi^2}\frac{\hat{m}^2 (s)}{s}
            + \left(\frac{8}{3}a_s(s) + \frac{59}{3}a_s^2(s)\right)
              \frac{\hat{m}\langle\bar{u}u+\bar{d}d\rangle}{s^2} 
            - \frac{16}{7\pi^2}\frac{\hat{m}^4(s)}{s^2} \nonumber \\
       & &  -\, 8 \pi^2a_s(\mu^2)\left[1 + \left(\frac{119}{24} - \frac{1}{2}
                                         L(s)\right) a_s(\mu^2)
                              \right]\frac{\langle {\cal O}_6^{1}(\mu^2)
                                      \rangle}{s^3} \nonumber \\
       & &  +\, \frac{2 \pi^2}{3}\left(3 + 4 \, L(s) \right)a^2_s(\mu^2)
              \frac{\langle {\cal O}_6^2(\mu^2)\rangle}{s^3}
              + \frac{\langle {\cal O}_8\rangle}{s^4}~, \\[0.2cm]
\label{eq_corr0}
     \Pi_{\bar u d,V-A}^{(0)}(-s) 
       &=& 
            -\, \frac{3}{\pi^2}\left[2a_s^{-1}(s)-5 + 
              \left( -\frac{21373}{2448} + \frac{75}{34} \zeta (3) 
              \right) a_s (s) \right]\frac{\hat{m}^2(s)}{s}
            - 4\hat{C}(\mu^2)\frac{\hat m^2(\mu^2)}{s}
             \nonumber \\
       & &
            -\, 2\frac{\hat{m}\langle\bar{u}u+\bar{d}d\rangle}{s^2} 
            - \frac{1}{7\pi^2}\left(\frac{53}{2} - 12a_s^{-1}(s)
              \right)\frac{\hat{m}^4(s)}{s^2}~,
\eeqn
with $a_s(s)=\alpha_s(s)/\pi$, $L(s)={\rm log}(s/\mu^2)$ and the
dimension $D=6$ operators
\beqn
  {\cal O}_6^1 
        &\equiv& \bar{u}\gamma_{\mu} \gamma_5 T^a d \bar d \gamma^{\mu}
                  \gamma_5 T^a u - \bar{u}\gamma_{\mu}  T^a d \bar d 
                  \gamma^{\mu} T^a u \nonumber \\
\label{eq_dim6}
  {\cal O}_6^2 
        &\equiv& \bar{u}\gamma_\mu\bar{d}d\gamma^\mu u
                 -\bar{u}\gamma_\mu\gamma_5d\bar{d}\gamma^\mu\gamma_5u,
\eeqn
where the $SU(3)$ generators $T^a$ are normalized so that 
${\rm tr}(T^a T^b )=\delta^{ab}/2$.
We use the average mass $\hat{m}\equiv(m_u + m_d)/2$ in the above
equations, \ie, we assume $SU(2)$ symmetry. The constant $\hat{C}(\mu^2)$
depends on the renormalization procedure\footnote
{
We will assume a renormalization scheme that preserves chiral symmetry, so
that $\hat C$ is the same for the vector and  axial correlators.
}
and should not af\/fect physical observables. The dimension 
$D=0$ contribution is of pure perturbative origin and is degenerate 
in all-orders of perturbation theory for vector and axial-vector 
currents. Dimension $D=2$ mass terms are calculated perturbatively to 
order $\alpha_s^2$ which suf\/f\/ices for the light $u,d,$ quarks. The 
coef\/f\/icient functions of the dimension $D=4$ operators for vector
and axial-vector currents have been calculated to subleading order in 
Refs.~\cite{generalis,chetyrkin2}. Their vacuum expectation values are 
expressed in terms of the scale invariant gluon and quark condensates.
Since the Wilson coef\/f\/icients of the gluon condensate are symmetric 
for vector and axial-vector currents, they vanish in the dif\/ference.
The expectation values of the dimension $D=6$ operators~(\ref{eq_dim6})
obey the inequalities $\langle {\cal O}_6^1(\mu^2)\rangle\ge0$ 
and $\langle {\cal O}_6^2(\mu^2)\rangle\le0$, which can be derived 
from f\/irst principles. The corresponding coef\/f\/icient functions 
were calculated by the authors of Refs.~\cite{lanin} in the chiral 
limit for which the $J=0$ contribution vanishes. For the 
dimension $D=8$ operators no such calculations are available in the 
literature, and we will assume that there is no logarithmic $s$ 
dependence in leading order $\alpha_s$. Again, the $J=0$ 
contribution vanishes in the chiral limit.
\vs
As constraints on the nonperturbative phenomenological operators 
introduced in Eqs.~(\ref{eq_corr10}) and (\ref{eq_corr0}) from 
theory alone are scarce, we will benef\/it from the information 
provided by the $(V-A)$ spectral moments in order to determine 
the magnitude of the OPE power terms at $M_\tau^2$. We therefore 
perform a combined f\/it of the IMSR (\ie, $l=-1$) which determines 
$L_{10}$, and the $l\ge0$ moments which adjust the nonperturbative 
contributions.
%
%
\section{Chiral Perturbation Theory}

The non-strange correlators~(\ref{eq_correlator}) have been calculated at
one-loop  level~\cite{gale1,gale2} and, most recently, at two-loop
level~\cite{kago1,kago2} 
in Standard $\chi$PT. In this paper we stick to the $O(p^4)$ one-loop order
for the following two reasons: $i)$~the high energy corrections are often
more important than the $O(p^6)$ chiral corrections (whose precise estimate
has not yet been fully completed~\cite{kagodmo}) and $ii)$~it is important to
proceed in the combined analysis order by order in quark masses. On the
other hand, we use the generalized version of $\chi$PT
(G$\chi$PT)~\cite{gchpt}, which allows us to investigate the sensitivity of
the analysis to the variation of the quark condensate and of  the quark mass
ratio $r=m_s/\hat m$. The standard $\chi$PT assumes~\cite{weinberg77} 
$2\hat m\langle\bar qq\rangle\simeq-F_{\pi}^2 M_{\pi}^2$ and $r\simeq 2M_K^2
/M_{\pi}^2-1\approx25.9$, whereas G$\chi$PT admits lower values of these 
two quantities~\cite{sterngchpt,gchpt}. It is interesting to investigate
whether the ALEPH \sf\ data are precise enough to have any impact on the
on-going debate about the size of $\langle \bar q q \rangle$. Anyhow, the
alterations of the standard $O(p^4)$ results for non-strange
correlators~(\ref{eq_correlator}) introduced by G$\chi$PT are marginal. They
merely concern the symmetry breaking $J=0$ component of the \sfs\ and most
of them are actually absorbed into the renormalization of $F_{\pi}$ $(F_K)$.
\vs
In order to make our analysis as independent of a particular truncation of
the $\chi$PT series as possible, we proceed in two steps. First, one def\/ines
a phenomenological quantity called $L_{10}^{\rm{ef\/f}}$ via the
contribution of the small circle~$C_2$ (see Fig.~\ref{fig_contour}) to the
integral~(\ref{eq_momtheorycontour}) of the chiral combination $V-A$ for
$l=-1$. $L_{10}^{\rm{ef\/f}}$ is then determined in the combined f\/it of
the IMSR and $l\geq 0$ moments. The result of this f\/it is independent of 
the $\chi$PT renormalization scale $\mu_{\chi{\rm{PT}}}$. The latter is 
used in the next step in order to relate $L_{10}^{\rm{ef\/f}}$ to the 
quark-mass independent, scale dependent constant 
$L_{10}^r(\mu_{\chi{\rm{PT}}})$ and f\/inally to other observables (from
$\pi\rightarrow e \nu \gamma$ data, $\langle r^2\rangle_{\pi}$).
%
%
\subsection{Non-strange IMSR's}

For the Cabibbo-allowed channel we obtain:
\beqn
\label{chpt1}
   \Pi^{(0+1)}_{\bar u d,V}(s)       &=& 
          4M^r_{KK}(s) + 8 M^r_{\pi\pi}(s) - 4(L_{10}^r + 2H_1^r)~,  \\
   \Pi^{(0)}_{\bar u d,V} (s)   &=& 0~, \\
   \Pi_{\bar u d,A}^{(0+1)} (s) &=& 
          - \frac{2F_\pi^2}{s-M_{\pi}^2} - 4(2H_1^r-L_{10}^r)~, \\
   s\Pi_{\bar u d,A}^{(0)} (s)  &=& 
          - \frac{2F_\pi^2M_{\pi}^2}{s-M_{\pi}^2} + 8{\hat m}^2(H_{2,2}-2B_3)~,
\eeqn
while, for the strange channel the correlators read
\beqn
\label{chpt2}
   \Pi^{(0+1)}_{\bar u s,V} (s)     &=& 
          6M^r_{K\eta}(s) + 6M^r_{K\pi}(s) - 4(L_{10}^r+2H_1^r )~,  \\
   s\Pi_{\bar u s,V}^{(0)}(s)       &=& 
          6\left(L_{K\eta}(s)+L_{K\pi}(s)\right) 
          + 2\hat{m}^2 (r-1)^2(2B_3+H_{2,2})~, \\ 
   \Pi_{\bar u s,A}^{(0+1)}(s)      &=& 
          - \frac{2 F_K^2}{s-M_{K}^2} - 4( 2H_1^r-L_{10}^r )~,  \\
   s\Pi_{\bar u s,A}^{(0)}(s)  &=& 
          - \frac{2F_K^2M_{K}^2}{s-M_{K}^2}
          + 2{\hat m}^2(r+1)^2(H_{2,2}-2B_3)~,
\eeqn
The functions $M^r_{PP^\prime}(s)$ and $L_{PP^\prime}(s)$ are loop 
integrals, def\/ined, \eg, in Ref.~\cite{gale2}. The superscript $r$ 
refers to renormalized quantities, which depend on the scale 
$\mu_{\chi{\rm{PT}}}$. The whole expressions are 
$\mu_{\chi{\rm{PT}}}$ independent. $H_{2,2}$ and 
$B_3$ are found to be f\/inite, in agreement with~\cite{marc}, and do not
need renormalization. $H_1^r$ and $H_{2,2}$ are coef\/f\/icients of
contact terms of the sources. They are counterterms needed to renormalize
the ultraviolet divergences of the Green functions and do not appear
in physical observables.  Our aim is to determine $L_{10}$:
therefore  we will  consider the dif\/ference between the vector and
the axial-vector correlators for which the constant $H_1$ disappears.
Correspondingly, as already pointed out, we will not need the
perturbative expressions which are identical for vector and
axial-vector cases.  As for the constant $H_{2,2}$ which multiplies
the term 
\[ 
   \langle D_{\mu} \chi^{\dagger} D^{\mu} \chi \rangle 
\]
of the ${\cal L}_{(2,2)}$ chiral lagrangian\footnote
{
${\cal L}_{(n,m)}$ collects terms in the chiral lagrangian with $n$~covariant
derivatives and $m$~powers of quark masses. In the same notation the $H_1$
constant introduced by Gasser and Leutwyler~\cite{gale2} would become
$H_{4,0}$. 
}
, it always 
appears in the same combination with $\hat C(\mu^2)$, in such a way 
that the ambiguities cancel out. We thus def\/ine a $\hat{H}_{2,2}$, in 
which the constant $\hat{C}(M_\tau^2)$ is absorbed. What is new at 
this order with respect to S$\chi$PT is the appearance of the  constant 
$B_3$ which multiplies the term 
\[ 
   \langle U^{\dagger}D_{\mu}\chi U^{\dagger}D_{\mu}\chi+h.c.\rangle 
\] 
of the ${\cal L}_{(2,2)}$ lagrangian.  As can be seen from its form
it is dif\/f\/icult to f\/ind a process in which $B_3$ would contribute 
directly. It will contribute to of\/f-shell vertices involving Goldstone 
bosons. 
\vs
The non-strange IMSR's corresponding to $l=-1$ and $k=0,1$ read:
\beqn
\label{eq_imsr0}
  \lefteqn{\frac{1}{|V_{ud}|^2S_{\rm EW}} R_{\tau,V-A}^{(0,-1)}} \nonumber \\ 
     &=&
         -\, 96 \pi^2 L_{10}^{\rm{ef\/f}} + 24 \pi^2 \frac{ F_{\pi}^2
	 M_{\pi}^2}{M_{\tau}^4} \nonumber \\
     & & 
         +\, \frac{144}{M_{\tau}^2}  \hat m^2 (M_{\tau}^2)   
             \left[\frac{1}{a_s(M_{\tau}^2)} - \frac{23}{8} + 
                   \left( \frac{\pi^2}{12} -\frac{36061}{4896} +
		   \frac{75}{34} \zeta (3)  \right) a_s(M_{\tau}^2)  
             \right] \nonumber \\
     & & 
         +\, \frac{96}{M_{\tau}^4} \pi^2  \hat m \langle  
           \bar d d + \bar u u \rangle 
           \left[ 1 + a_s(M_{\tau}^2)  + \frac{17}{2} a_s^2(M_{\tau}^2) 
           \right] -\ \frac{576}{7 M_{\tau}^4}  \hat m^4 (M_{\tau}^2)  
           \left[ \frac{1}{a_s(M_{\tau}^2)} - \frac{29}{24}
           \right] \nonumber  \\
     & & 
         -\, \frac{ 192 \pi^4}{M_{\tau}^6} a_s (\mu^2) 
             \left[ 1 + \left(\frac{103}{24} 
                              -\frac{1}{2}L(M_\tau^2)\right)a_s(\mu^2)
             \right] \langle {\cal O}_6^1(\mu^2) \rangle  \nonumber \\
     & &
         +\, \frac{ 400 \pi^4}{3M_{\tau}^6}
	     a_s^2(\mu^2)\left(1 + \frac{12}{25}L(M_\tau^2)\right)  
             \langle {\cal O}_6^2(\mu^2) \rangle~,
\eeqn
\beqn
\label{eq_imsr1}
  \lefteqn{\frac{1}{|V_{ud}|^2S_{\rm EW}} R_{\tau,V-A}^{(1,-1)}} \nonumber \\
     &=& 
	 -\, 96 \pi^2 L_{10}^{\rm{ef\/f}} - 24 \pi^2 \left( \frac{
	 F_{\pi}^2}{M_{\tau}^2} - 3 \frac{ F_{\pi}^2 M_{\pi}^2}{M_{\tau}^4}
	 + \frac{F_{\pi}^2 M_{\pi}^4}{M_{\tau}^6} \right) \nonumber \\
     & & 
         +\, \frac{144}{M_{\tau}^2}  \hat m^2(M_{\tau}^2 )  
             \left[\frac{1}{a_s(M_{\tau}^2)} - \frac{71}{24}  + 
                   \left( \frac{ \pi^2}{12}  - \frac{39461}{4896} +
		   \frac{75}{34} \zeta (3) \right) a_s(M_{\tau}^2) 
             \right] \nonumber \\
     & & 
         +\, \frac{144}{M_{\tau}^4} \pi^2\hat m\langle\bar dd+\bar uu\rangle
             \left[ 1 + \frac{2}{3} a_s(M_{\tau}^2)
                    +\frac{43}{6} a_s^2 (M_{\tau}^2)
             \right] - \frac{864}{7 M_{\tau}^4}\hat m^4(M_{\tau}^2)
             \left[\frac{1}{a_s(M_{\tau}^2)} - \frac{2}{3}
             \right] \nonumber \\
     & & 
         -\, \frac{ 480 \pi^4}{M_{\tau}^6} a_s (\mu^2) 
             \left[ 1 + \left(\frac{581}{120} 
                              - \frac{1}{2}L(M_\tau^2)\right)a_s(\mu^2) 
             \right]\langle {\cal O}_6^1(\mu^2) \rangle  \nonumber \\
     & &
         +\, \frac{472 \pi^4}{3M_{\tau}^6}a_s^2(\mu^2)
	     \left(1 + \frac{60}{59}L(M_\tau^2)\right)
             \langle {\cal O}_6^2(\mu^2) \rangle 
             + 24\pi^2\frac{\langle {\cal O}_8\rangle}{M_\tau^8}~,
\eeqn
where we have def\/ined
\beqn \label{eq_defl10eff} 
-8 L_{10}^{\rm{ef\/f}} 
	&=&
	     \lim_{s \to 0} 
             \left\{ \left(1+\frac{2s}{M_{\tau}^2} \right) 
                     \Pi_{\bar u d,{\rm{V-A}}}^{(0+1)}(s)
                         -\frac{2 s}{M_{\tau}^2} 
                     \Pi_{\bar u d, {\rm{V-A}}}^{(0)}(s)
                         -\frac{2 F_{\pi}^2}{s-M_{\pi}^2} 
                     - 4 \frac{ F_{\pi}^2}{M_{\tau}^2} 
             \right\} \nonumber \\
	& &	
              +\, 8 \frac{ \hat m^2(M_{\tau}^2)}{M_{\tau}^2} 
                    \hat C(M_{\tau}^2)~,
\eeqn
which is proportional to the contribution of the small circle $C_2$ 
to the integral~(\ref{eq_momtheorycontour}), with the pion pole  
subtracted. This quantity is a well def\/ined observable, the 
ambiguity in the two-point function being absorbed by the constant 
$\hat C$. In the particular case of the one-loop G$\chi$PT calculation, 
its expansion reads: 
\beq \label{eq_l10eff}
   L_{10}^{\rm{ef\/f}} 
        = L_{10}^r(\mu_{\chi{\rm{PT}}}) + \frac{1}{128 \pi^2} 
          \left( \log\frac{M_{\pi}^2}{\mu_{\chi{\rm{PT}}}^2}+1 \right) 
          + \frac{1}{384 \pi^2} \log\frac{M_K^2}{M_{\pi}^2} 
          + \frac{2\hat m^2}{M_{\tau}^2}\left(2B_3-\hat H_{2,2}\right)~,
\eeq
which is independent of $\mu_{\chi{\rm{PT}}}$. Unless stated otherwise
all condensates, quark masses and $\chi$PT constants in the above 
expressions are evaluated at QCD renormalization scale 
$\mu_{\rm QCD}=M_\tau$, while the product of 
the light quark mass and the scalar quark operator, 
$\hat{m}\langle\bar dd+ \bar uu\rangle$, is scale invariant.
Taking the dif\/ference of Eq.~(\ref{eq_imsr0}) 
and (\ref{eq_imsr1}) and subtracting the contribution from the pion 
pole recovers the expression for $R_{\tau,V-A}=R_{\tau,V-A}^{(0,0)}$ 
given in~\cite{bnp}. Due to the strong intrinsic
correlations of $98\%$ between the IMSR's def\/ined above only one 
IMSR is used as input to the combined f\/it. We f\/ind it convenient 
to use the moment $k=1,l=-1$ (Eq.~(\ref{eq_imsr1})) because its 
experimental value is known with a $30\%$ better precision which 
is due to the additional $(1-s/M_\tau^2)$ suppression of the less 
accurate high energy tail of the $(V-A)$ \sf.
%
%
\subsection{Strange IMSR}

Analogous IMSR's with $l=-2$ would require the two-loop results for 
the correlators~(\ref{eq_correlator}), because quark-mass independent terms
from the  
${\cal L}_{(6,0)}$ lagrangian would give rise to new contributions which are
not {\it a priori} small. However, if we 
consider the dif\/ference between the strange and non-strange $(V+A)$  
moments these terms cancel out because of the $SU(3)$ symmetry. Of course
there will  
be the two-loop corrections to the terms already present in the 
one-loop results, but these are subleading. Therefore, we can write 
a particular combination of inverse moments which does not contain any
unknown low-energy constant. An example of such a combination is
$4 R_{\tau,V+A}^{(1,-1)} + R_{\tau,V+A}^{(2,-2)}$, so that the strange 
IMSR takes the form: 
\beqn
\label{eq_strimsr}
 	\lefteqn{
           \frac{1}{S_{\rm{EW}}} \left[ \frac{1}{|V_{us}|^2} 
           \left( 4R^{1,-1}_{\tau,S} + R^{2,-2}_{\tau,S} 
           \right) - \frac{1}{|V_{ud}|^2} 
           \left( 4R^{1,-1}_{\tau,V+A} + R^{2,-2}_{\tau,V+A}
           \right)            \right]} \nonumber \\
 &=& 
 	12 \pi^2 \Bigg\{ 4 M_{\tau}^2 \left[ \frac{3}{2} \left( M^{r\prime}_{K
	\eta}(0) + M^{r\prime}_{K\pi}(0) \right) -  {M^{r\prime}_{KK}}(0) - 2
	{M^{r\prime}_{\pi\pi}}(0)  \right] \nonumber \\
 & &
       +\, 12 \left( M^r_{K\eta}(0) + M^r_{K \pi}(0) \right)- 8 \left(
	M^r_{KK}(0) +  2 M^r_{\pi\pi}(0) \right) - 12 \left(
	L^\prime_{K\eta}(0) + L^\prime_{K \pi}(0) \right) \nonumber \\
 & & 
        +\, \frac{12}{M^2_{\tau}} \left( F^2_{K} - F^2_{\pi} \right) -
	\frac{16}{M^4_{\tau}} \left( F^2_{K} M_{K}^2 - F_{\pi}^2 M_{\pi}^2
	\right) + 
	\frac{6}{M_{\tau}^6} \left( F_{K}^2 M_{K}^4 - F_{\pi}^2 M_{\pi}^4
	\right) 
	\Bigg\}_{\approx -7.91 } \nonumber \\
 & &
 	+\, 180 \frac{ \hat m^2}{M_{\tau}^2} (r^2 - 1) \left[ 1 +
	\frac{59}{15}  
	a_s(M_{\tau}^2) \right] \nonumber \\
 & & 
 	-\, 192 \pi^2 \frac{1}{M_{\tau}^4} \langle  m_s \bar
	s s -\hat m \bar u u  \rangle \left[ 1 + \frac{1}{2}
	a_s(M_{\tau}^2)  + \frac{151}{24} a_s^2(M_{\tau}^2) \right]
	\nonumber \\ 
 & & 
 	+\, \frac{576}{7} \frac{\hat m^4}{M_{\tau}^4} \left[ (r^4 - 1)
 	\frac{1}{a_s(M_{\tau}^2)} + \frac{215}{48} - \frac{7}{2} r^2
	-\frac{47}{48} r^4 \right] \nonumber \\       
 & &
	+\, \frac{156\pi^2}{M_{\tau}^6} \langle {\cal O}_6^{(\Delta S)} \rangle +
	\frac{72 \pi^2}{M_{\tau}^8} \langle {\cal O}_8^{(\Delta S)} \rangle~,
\end{eqnarray}
where $-7.91$ is the value of the $\chi$PT contribution, using the $\pi^0$
mass and the QCD $K^+$ mass (\ie, without electromagnetism).
We have neglected all logarithmic $s$ dependence in the dimension $D=6$
and $D=8$ operators ${\cal O}_6^{(\Delta S)}$ and  ${\cal O}_8^{(\Delta
S)}$. The latter are expected to be suppressed because, in contrast to the
non-strange ${\cal O}_6$ and ${\cal O}_8$, they vanish in the chiral limit. 
Data for the inclusive vector plus axial-vector strange \sf\ from $\tau$ 
decays are not available at present. Such data could provide information 
on the size of the quark condensate, which could represent up to $10\%$ 
of the $\chi$PT contribution to the r.h.s. of Eq.~(\ref{eq_strimsr}).
%
%
\section{Theoretical parameters and uncertainties}
\label{sec_theoerr}

When f\/itting the theoretical prediction of the $R_{\tau,V-A}^{(k,l)}$ 
moments to data, theoretical as well as experimental uncertainties
and the correlations of these between the $(k,l)$ moments must be
considered. The masses of the light quarks are parametrized using
the mass ratio $r=m_s/\hat{m}$ of which the central value is set 
to the S$\chi$PT value of 26. A lower limit is found at
$r\ge r_{\rm limit}=2(M_K/M_\pi)-1\approx6.1$ (while 
$r_{\rm limit}\approx8.2$ when including higher orders~\cite{sternmainz})
which determines the range
\beqns
 8 < r < \infty~.
\eeqns
The average light quark mass is then obtained \via\ $\hat{m}=m_s/r$
where we use for the strange quark mass
$m_s(M_\tau)=172~{\rm MeV}/c^2$~\cite{shaomin}.
This parametrization makes it possible to use the theoretical correlation
between $\hat m$ and the quark condensate, which to leading order in quark
masses is given by the generalized Gell-Mann-Oakes-Renner
relation~\cite{sterngchpt,gchpt}:
\beq
\label{eq_ggmor}
   \hat{m}\langle \bar{u}u + \bar{d}d\rangle
       \simeq - F_\pi^2 M_\pi^2\frac{(r - r_1)(r + r_1 + 2)}{r^2 - 1}~,
\eeq
where $r_1\simeq2(M_K/M_\pi)-1$. For the standard value
$r=25.9$, Eq.~(\ref{eq_ggmor}) becomes the usual PCAC relation
$\hat{m}\langle\bar{u}u+\bar{d}d\rangle=-F_\pi^2 M_\pi^2$.
Corrections to Eq.~(\ref{eq_ggmor}) are expected to be small in the 
whole range of~$r$ so that we assume a relative uncertainty of $10\%$.
We will comment in Section~\ref{sec_results} on the sensitivity of 
the data with respect to the $r$ ratio. Theoretical uncertainties are 
introduced from the strong coupling constant where, in order to be 
uncorrelated to the $\tau$ data used in this analysis, we rely on 
the result from the global electroweak f\/it found recently to 
be~\cite{jerusalem,alphapap}
\beqns
   \alpha_s(M_Z^2) = 0.1198 \pm 0.0031~.
\eeqns

Uncertainties from the OPE separation scale $\mu$ are evaluated
by varying $\mu$ from 1.3~GeV to 2.3~GeV, while in the f\/it we 
choose $\mu=M_\tau$ so that the logarithmic scale dependence of the 
dimension $D=6$ terms vanishes after the contour integration. 
Additional small uncertainties stem from the 
pion decay constant, $F_\pi=(92.4\pm0.2)~{\rm MeV}$, taken from 
Ref.~\cite{pdg} and the overall correction factor for electroweak 
radiation, $S_{\rm EW}=1.0194$, obtained in Ref.~\cite{marciano},
with an estimated error of $\Delta S_{\rm EW}=0.0040$ according to 
Ref.~\cite{g_2pap}. 
\vs
An overview of the associated uncertainties in the theoretical 
prediction of the moments is given in Table~\ref{tab_momdata}.
The moment errors from the $\alpha_s$ uncertainty depend 
on the central input values of the nonperturbative operators. The
numbers given in the fourth line of Table~\ref{tab_momdata}
correspond to the f\/it values, Eqs.~(\ref{eq_o6})--(\ref{eq_o8}),
which have been obtained in an iterative procedure.
%
%
\section{\SFs\ from hadronic \boldmath$\tau$ decays}

\begin{figure}[t]
\epsfxsize12.cm 
\centerline{\epsffile{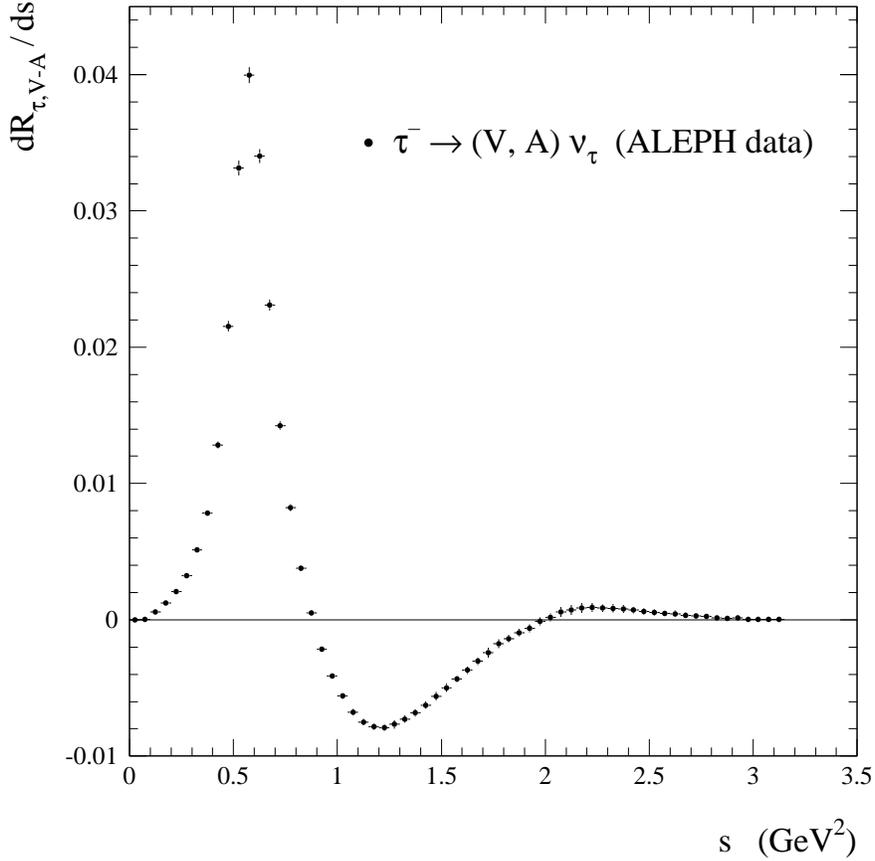}}
  \caption[.]{\label{fig_massvma}\it
           Vector minus axial-vector $(V-A)$ invariant mass-squared 
           distribution measured by ALEPH~\rm\cite{aleph_asf}.}
\end{figure}

The ALEPH Collaboration measured the inclusive invariant mass-squared
spectra of vector and axial-vector hadronic $\tau$ decays and
provided the corresponding bin-to-bin covariance 
matrices~\cite{aleph_vsf,aleph_asf}. The mass distributions
naturally contain the kinematic factor of Eq.~(\ref{eq_momtheory}) 
so that the measured spectral moments read
\beq
\label{eq_momdata}
   R_{\tau,V-A}^{(k,l)} = 
       \intl_0^{M_\tau^2} ds\,\left(1-\frac{s}{M_\tau^2}\right)^{\!\!k}
                              \left(\frac{s}{M_\tau^2}\right)^{\!\!l}
       \left[B_V\frac{dN_V}{N_V\,ds}-B_A\frac{dN_A}{N_A\,ds}\right]
       \frac{1}{B_e}~,
\eeq
with the normalized invariant mass-squared spectra 
$(1/N_{V/A})(d N_{V/A}/d s)$ of vector and axial-vector f\/inal 
states, the electronic \br\ (using universality)~\cite{pdg,aleph_asf}, 
$B_e=(17.794\pm0.045)\%$, and the inclusive \brs~\cite{aleph_asf},
$B_V=(31.58\pm0.29)\%$, $B_A=(30.56\pm0.30)\%$, as well as their 
dif\/ference, $B_{V-A}=(1.02\pm0.58)\%$. Due to anticorrelations 
between vector and axial-vector f\/inal states, especially for 
the ${\rm K}\bar{\rm K}\pi$ modes where the vector and axial-vector 
parts are unknown, the error of the dif\/ference is larger than 
the quadratic sum of the errors on $V$ and $A$. 
F\/igure~\ref{fig_massvma} shows the $(V-A)$ mass-squared distribution,
which is the integrand of Eq.~(\ref{eq_momdata}) for zero moments, 
$k=l=0$. With increasing masses it is dominated by the $\rho$ $(V)$, 
$a_1$ $(A)$ and the $\rho(1450)$, $\omega\pi$ $(V)$ resonance 
contributions which create the oscillating behaviour. 
Table~\ref{tab_momdata} 
and \ref{tab_momcorr} give the experimental values and uncertainties 
for the IMSR $R_{\tau,V-A}$ and the $k=1$, $l=0,\dots,3$ moments 
as well as their correlations which are computed analytically from 
the contraction of the derivatives of the moments with the covariance 
matrices of the respective normalized invariant mass-squared 
spectra.
\vs
\begin{figure}[t]
  \epsfxsize12.cm
  \centerline{\epsffile{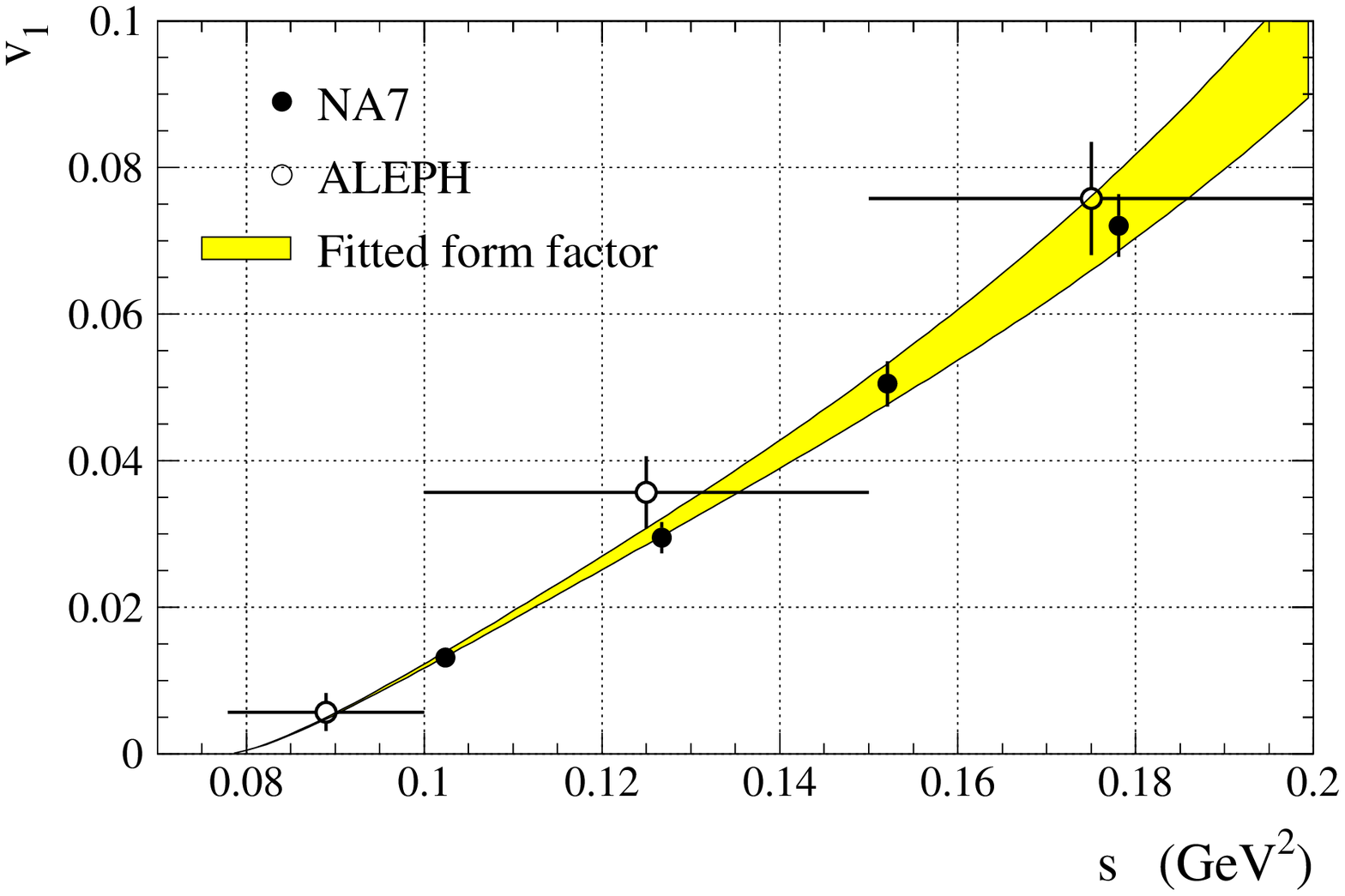}}
  \caption[.]{\label{fig_lowen}\it
              Low energy vector \sfs\ from $\tau$ decays and, \via\ 
              CVC, from $e^+e^-\rightarrow\pi^+\pi^-$ data measured 
              by NA7~\rm\cite{na7}.}
\end{figure}
Based on isospin invariance, the conserved vector current hypothesis 
(CVC) relates vector hadronic $\tau$ \sfs\ to isovector cross section 
measurements of the reaction $e^+e^-\rightarrow\,{\rm hadrons}$. There 
exist precise data on the low energy, time-like pion form 
factor-squared $|F_\pi(s)|^2$ measured by the NA7 
Collaboration~\cite{na7}. Using the CVC relation 
\beq
\label{eq_ffv}
   v_{1,\,\pi^-\pi^0}(s) = 
             \frac{1}{12}\left(1-\frac{4M_\pi^2}{s}\right)^{\!\!3/2} 
             |F^{I=1}_\pi(s)|^2~,
\eeq
one can include the additional data in order to improve the precision
of the moments~(\ref{eq_momdata}), in particular for the IMSR
in which the low-energy region is emphasized. 
Figure~\ref{fig_lowen} shows the vector \sf\ from $\tau$ data 
(three bins) together with the NA7 measurements for energy-squared 
$s\le0.2~{\rm GeV}^2$. In addition, we give  the result when 
f\/itting both data sets using the parametrization
\beq
\label{eq_fffit}
    F_\pi(s) = 1 + \frac{1}{6}\langle r^2 \rangle_\pi s + As^2 + Bs^3~,
\eeq
for the pion form factor. Using analyticity, the pion charge 
radius-squared, $\langle r^2 \rangle_\pi=(0.439\pm0.008)~{\rm fm}^2$, 
is taken from an analysis of space-like data~\cite{space_like}.
We obtain the f\/it results $A=-(7.5\pm1.1)~{\rm GeV}^{-4}$ 
and $B=(62.5\pm6.4)~{\rm GeV}^{-4}$ with $\chi^2=0.6$ for 5 degrees 
of freedom. The correlation between $A$ and $B$ is absorbed in the 
diagonal errors given, so that both quantities can be handled as 
being uncorrelated. Replacing for the above energy interval 
$4M_\pi^2\le s\le0.2~{\rm GeV}^2$ the pure $\tau$ data by a 
combination of $\tau$ and $e^+e^-$ data represented by
the analytical expressions~(\ref{eq_ffv}) and (\ref{eq_fffit}),
we obtain the results given in the third and fourth line of 
Table~\ref{tab_momdata}. A small improvement in precision of $11\%$ 
is observed for the IMSR.
\vs
The spectral information is used to f\/it simultaneously the
low-energy quantity $L_{10}^{\rm{ef\/f}}$ and the nonperturbative 
phenomenological operators. For dimension $D=6$ we will neglect the 
contribution of ${\cal O}_6^2$, which is suppressed by $\alpha_s^2$ 
and, furthermore, is suppressed relatively to ${\cal O}_6^1$ in the 
large~$N_c$ limit. Therefore we will simply keep 
${\cal O}_6={\cal O}_6^1(M_\tau^2)$ and the ${\cal O}_8$ operator 
of dimension $D=8$.
%
%
\section{Results of the f\/it}
\label{sec_results}	

\begin{figure}[t]
  \epsfxsize12.cm
  \centerline{\epsffile{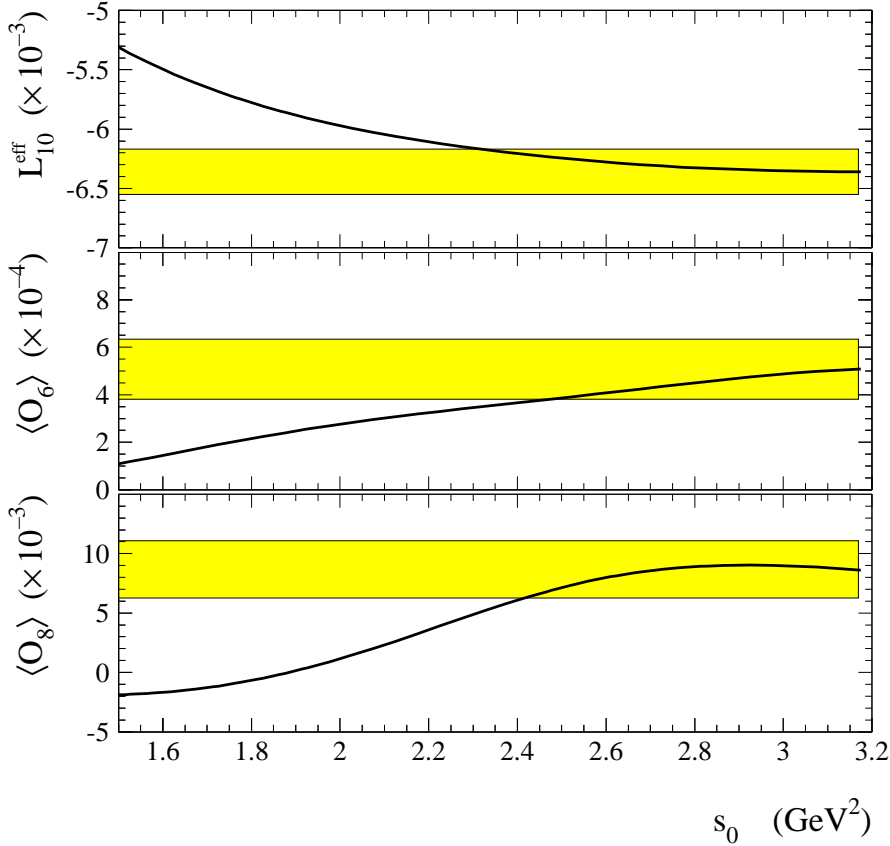}}
  \caption[.]{\label{fig_running}\it
              Fit results for $L_{10}^{\rm ef\/f}$ and the 
              nonperturbative operators as a function of the 
              ``$\tau$ mass'' $s_0$. The bands depict the 
              values~\rm(\ref{eq_l10})--(\ref{eq_o8}) \it within 
              errors, obtained at $M_\tau^2$.}
\end{figure}

The f\/it minimizes the $\chi^2$ of the dif\/ferences between measured
and f\/itted quantities contracted with the inverse of the sum of the
experimental and theoretical covariance matrices taken from
Table~\ref{tab_momcorr}.
The results of the f\/it are for $L_{10}^{\rm{ef\/f}}$:
\beq
   \begin{array}{|rcl|}\hline 
        & & \\
  \label{eq_l10}
   ~~~L_{10}^{\rm{ef\/f}}  
        &=& -(6.36\pm0.09_{\rm exp}\pm0.14_{\rm theo}
             \pm0.07_{\rm f\/it} \pm0.06_{\rm OPE}) \times 10^{-3}~,~~~ \\
        & & \\ \hline
  \end{array} 
\eeq
and for the nonperturbative operators:
\beqn
  \label{eq_o6}
   \langle {\cal O}_6 \rangle 
        &=&  (5.0\pm0.5_{\rm exp}\pm0.4_{\rm theo}
             \pm0.2_{\rm f\/it}\pm1.1_{\rm OPE}) 
             \times 10^{-4}~{\rm GeV}^6~, \\[0.3cm]
  \label{eq_o8}
   \langle {\cal O}_8\rangle          
        &=&  (8.7\pm1.0_{\rm exp}\pm0.1_{\rm theo}
             \pm0.6_{\rm f\/it}\pm2.1_{\rm OPE}) \times 10^{-3}~{\rm GeV}^8~,
\eeqn
\begin{table}[p]
  \begin{center}
{\small
  \begin{tabular}{|l||cccccc|} \hline 
   \mc{1}{|r||}{$(k,l)\longrightarrow$}  &
  $(1,-1)$ &  $(0,0)$ &  $(1,0)$ & $(1,1)$ &  $(1,2)$ & $(1,3)$ 
\\ \hline\hline
 $R_{\tau,V-A}^{(k,l)}$ (ALEPH) &
   $5.16$  & $0.055$  & $0.038$  & $0.047$ &$-0.0164$ & $-0.0126$ \\ 
 $\Delta^{\rm exp} R_{\tau,V-A}^{(k,l)}$  &
   $0.09$  & $0.031$  & $0.017$  & $0.006$ & $0.0035$ & $ 0.0023$ \\ 
\hline
 $R_{\tau,V-A}^{(k,l)}$ (ALEPH + NA7) &
   $5.13$  & $0.055$  & $0.037$  & $0.047$ &$-0.0164$ & $-0.0126$ \\ 
 $\Delta^{\rm exp} R_{\tau,V-A}^{(k,l)}$  &
   $0.08$  & $0.031$  & $0.017$  & $0.006$ & $0.0035$ & $ 0.0023$ \\ 
\hline\hline
 $\Delta^{\rm theo} R_{\tau,V-A}^{(k,l)}$ ($\Delta r$)  &    
   $0.12$  & $0.003$  & $0.003$  & $0.001$ & $0.0003$ & $<0.0001$ \\ 
 $\Delta^{\rm theo} R_{\tau,V-A}^{(k,l)}$ ($\Delta \alpha_s$) &    
   $0.02$  & $0.009$  & $0.009$  & $0.002$ & $0.0029$ & $0.0001$  \\ 
 $\Delta^{\rm theo} R_{\tau,V-A}^{(k,l)}$ ($\Delta S_{\rm EW}$) &    
   $0.02$  & $<0.001$ & $<0.001$ & $<0.001$& $ 0.0001$& $<0.0001$ \\ 
 $\Delta^{\rm theo} R_{\tau,V-A}^{(k,l)}$ ($\Delta \mu_{\rm OPE}$) &    
   $<0.01$ & $0.005$  & $0.005$  & $0.002$ & $ 0.0018$& $<0.0001$ \\ 
 $\Delta^{\rm theo} R_{\tau,V-A}^{(k,l)}$ ($\Delta \langle\bar{q}q\rangle$) &
   $<0.01$ & $<0.001$ & $<0.001$ & $<0.001$& $<0.0001$& $<0.0001$ \\ 
 $\Delta^{\rm theo} R_{\tau,V-A}^{(k,l)}$ ($\Delta F_\pi$) &    
   $<0.01$ & $<0.001$ & $<0.001$ & $<0.001$& $<0.0001$& $<0.0001$ \\ 
\hline\hline
$R_{\tau,V-A}^{(k,l)}$ (Theory f\/itted) &
   $5.13$  & $0.061$  & $0.032$  & $0.053$ & $-0.0148$& $-0.0098$ \\
\hline
  \end{tabular}
}
  \end{center}
  \caption[.]{\label{tab_momdata}\it
              Measured spectral Moments of vector $(V)$ minus axial-vector 
              $(A)$ using $\tau$ data only (ALEPH) and using $\tau+e^+e^-$
              data (ALEPH + NA7). 
              The quoted errors account for the total experimental 
              uncertainties including statistical and systematic effects
              as well as the theoretical uncertainties according to 
              Section~\rm\ref{sec_theoerr}\it. The last line gives the 
              fitted theoretical moments using the parameters
              given in Eqs.~\rm(\ref{eq_l10})--(\ref{eq_o8})\it.}
\vspace{0.5cm}
  \begin{center}
\begin{tabular}{|l||cccccc|} \hline
\mc{1}{|c||}{
 $(k,l)$} &$(1,-1)$&$(0,0)$&$(1,0)$&$(1,1)$&$(1,2)$&$(1,3)$ 
\\ \hline\hline 
 $(1,-1)$ & 1      & 0.46  & 0.61  & 0.40  & 0.26  & 0.13 \\
 $(0,0)$  & --     & 1     & 0.89  & 0.97  & 0.84  & 0.80 \\
 $(1,0)$  & --     & --    & 1     & 0.88  & 0.74  & 0.45 \\
 $(1,1)$  & --     & --    & --    & 1     & 0.89  & 0.78 \\
 $(1,2)$  & --     & --    & --    & --    & 1     & 0.76 \\ 
\hline
\end{tabular}
  \end{center}
  \caption[.]{\label{tab_momcorr}\it
              Sum of experimental and theoretical correlations between the
              moments $R_{\tau,V-A}^{(k,l)}$.}
\vspace{0.5cm}
  \begin{center}
  \begin{tabular}{|l||ccc|} \cline{2-4}
  \mc{1}{c|}{$ $}            & $L_{10}$ & $\langle {\cal O}_6\rangle$ 
                                                  & $\langle {\cal O}_8\rangle$    
\\ \cline{2-4}\hline
  $L_{10}$                   &  1       & $-0.26$ & $ 0.05$  \\
  $\langle{\cal O}_6\rangle$ & --       & $ 1   $ & $ 0.14$  \\ 
\hline
  \end{tabular}
  \end{center}
  \caption[.]{\label{tab_rescorr}\it
              Correlations between the fitted 
              parameters~\rm(\ref{eq_l10})--(\ref{eq_o8})\it.}
\end{table}
with a $\chi^2$ of 2.5 for 3 degree of freedom. The errors
are separated in experimental (f\/irst number) and theoretical
(second number) parts, and a f\/it uncertainty (third number) is 
added. The latter is due to a well known bias when f\/itting 
quantities for which correlations are due to normalization
uncertainties~\cite{agostini} (here the $\tau$ \brs) leading 
systematically to lower values in terms of the normalization of 
the f\/itted parametrization. The errors quoted account for the 
dif\/ferences between fully correlated and uncorrelated results. 
The authors of Ref.~\cite{aleph_asf} observed a variation of the 
results on the nonperturbative operators depending on the weighting 
of the $\tau$ \sfs\ used in the actual f\/it. These variations stem 
from deviations between data and the OPE approach for the running 
$R_{\tau,V/A}(s_0\le M_\tau^2)$ in the vector and axial-vector
channels (visualized in F\/ig.~17 of Ref.~\cite{aleph_asf}) and
from the corelation between the f\/itted dimension $D=6$ and $D=8$ 
operators. They have been found to be larger than the theoretical and 
experimental uncertainties. We repeat this study here in order 
to estimate the corresponding systematic uncertainties for the
f\/itted quantities. The last numbers in
Eqs.~(\ref{eq_l10})--(\ref{eq_o8}), denoted as ``OPE'' errors, 
give the deviations found. They are small for $L_{10}^{\rm{ef\/f}}$ 
and dominant for the nonperturbative operators.
\vs
Table~\ref{tab_rescorr} gives the correlations between the 
f\/itted parameters which are found to be small.
   Nevertheless, the interpretation of the parameter errors given 
   in Eqs.~(\ref{eq_l10})--(\ref{eq_o8}) as individual errors must
   be done with care in the presence of non-vanishing correlations.
   The results can reliably be used when applying the 
   whole expansion~(\ref{eq_ope}) which yields Eqs.~(\ref{eq_imsr0})
   and (\ref{eq_imsr1}).
\vs
Expressing $L_{10}^{\rm ef\/f}$ of Eq.~(\ref{eq_l10}) by means of
Eq.~(\ref{eq_l10eff}) at the $\chi$PT renormalization scale 
$\mu_{\chi\rm PT}=770~{\rm MeV}$, we obtain 
\beq
\label{eq_l10mrho}
   L_{10}^r(M_\rho)=-(5.13\pm0.19)\times10^{-3}~.
\eeq
In deriving the above value the term~$2\hat{m}^2(2B_3-\hat{H}_{2,2})$ 
in Eq.~(\ref{eq_l10eff}) has been neglected.  Na\"{\i}ve dimensional 
analysis estimates~\cite{georgi} give for the low-energy constants 
$B_3$ and $H_{2,2}$ an order of magnitude of $10^{-2}$, leading to a 
contribution which is negligible compared to the theoretical error in 
Eq.~(\ref{eq_l10}). The result~(\ref{eq_l10mrho}) is to be compared 
with the one-loop value of $L_{10}^r$ obtainable from 
$\pi \rightarrow e\nu\gamma$ decays and $\langle r^2\rangle_{\pi}$. 
The error in the latter has been decreased since the f\/irst 
determination of $L_{10}^r$ in Ref.~\cite{gale2}. Using the value  
$\langle r^2\rangle_{\pi}=(0.439\pm0.008)~{\rm fm}^2$~\cite{space_like}, 
one obtains 
\beq
\label{eq_l9rpi}
   L_9^r(M_{\rho})= ( 6.78 \pm 0.15 ) \times 10^{-3}~,
\eeq
which updates the value of Refs.~\cite{gale2,daphne}. The $L_{10}^r$ 
constant can be determined from the one-loop expression of the axial 
form factor, $F_A$, of $\pi \rightarrow e \nu \gamma$ (see
Ref.~\cite{pienugamma} for notations):
\beq
   F_A = \frac{4\sqrt{2}\,M_\pi}{F_{\pi}}(L_9+L_{10})~.
\eeq
Taking the value $F_A=0.0116\pm0.0016$~\cite{pdg}, one obtains
\beq
\label{eq_l9l10}
   L_9 + L_{10}  = (1.36 \pm 0.19 ) \times 10^{-3}~,
\eeq
and hence
\beq \label{eq_l10mrhogl}
   L_{10}^r(M_{\rho}) = ( -5.42 \pm 0.24) \times 10^{-3}~.
\eeq
This is the updated value of $L_{10}$ to be compared with our
result~(\ref{eq_l10mrho}). Note that the quoted errors in 
Eqs.~(\ref{eq_l10mrho})--(\ref{eq_l10mrhogl}) do not take into 
account uncertainties from higher order chiral corrections.
A two-loop evaluation for the combination 
$L_9+L_{10}$ has been attempted in Ref.~\cite{pienugamma}. However, 
this analysis is af\/fected by the controversial question about the 
value of the $SU(2)$ constant $l_2$. Using for example the values of
Ref.~\cite{l1l2} for the constants $l_1$ and $l_2$ on the basis of the 
two-loop precision $\pi\pi$ scattering phenomenology, we obtain 
$L_9 + L_{10} = (1.07 \pm 0.19 )\times 10^{-3}$ instead of 
$L_9 + L_{10}  = (1.57 \pm 0.15 )\times 10^{-3}$ given in 
Ref.~\cite{pienugamma}. 
\vs
The total, purely nonperturbative contribution to $R_{\tau,V-A}$ 
found in the f\/it, taking into account the correlations between
the operators, amounts to 
\beq
\label{eq_np}
      R_{\tau,V-A}  =  0.061 \pm 0.014~,
\eeq
compared to the measurement $R_{\tau,V-A}=0.055 \pm 0.031$. The 
reduced error of the theoretical f\/it to data compared to the 
measurement stems from the additional information used in the 
f\/it which is obtained from the shape of the \sfs\ and the OPE
constraint. The result~(\ref{eq_np}) is in good agreement with 
the value of $R_{\tau,V}-R_{\tau,A}=0.068$ found in 
Ref.~\cite{aleph_asf} . This is a non-trivial result
keeping in mind the logarithmic $s$ dependence of the dimension
$D=6$ Wilson coef\/f\/icients used in this analysis compared
to the vacuum saturation hypothesis adopted in Ref.~\cite{aleph_asf}.
In addition, in Ref.~\cite{aleph_asf}, vector and axial-vector were 
not combined in a simultaneous f\/it. The smaller systematic error
on the nonperturbative parts which is found in this analysis, in 
particular the reduced uncertainty from the explicit dependence of 
the moments employed, is due to the reduced correlation between the 
f\/itted $D=6$ and $D=8$ operators (see Table~\ref{tab_rescorr}).
The dimension $D=6$ contribution to $R_{\tau,V-A}$ corresponding 
to our f\/it result Eq.~(\ref{eq_o6}) amounts to 
$R_{\tau,V-A}^{(D=6)}=0.071\pm0.018$, which is signif\/icantly
than what one obtains from the vacuum saturation hypothesis~\cite{bnp},
$R_{\tau,V-A}^{(D=6)}\simeq0.97\times256\pi^3\alpha_s
\langle\bar qq\rangle^2/M_\tau^6\approx0.012$.
\vs
In addition to the test of the OPE by varying the $(k,l)$ moments 
used to f\/it $L_{10}^{\rm ef\/f}$ and the nonperturbative 
operators, we perform f\/its for variable ``$\tau$ masses'' 
$s_0\le M_\tau^2$~\cite{aleph_asf} which provides a direct 
test of the parameter stability at $M_\tau^2$. In order to 
perform such a study one has to replace all $\tau$ masses 
in Eqs.~(\ref{eq_momtheorycontour}), (\ref{eq_imsr1}) and 
(\ref{eq_momdata}) by $s_0$, while the latter must be corrected
by the kinematical factor $(1-s/s_0)(1+2s/s_0)/s_0$.
The scale invariance of the dimension $D=6$ operator for 
variable $s_0$ is approximately conserved when keeping the
scale parameter $\mu=M_\tau$ in Eqs.~(\ref{eq_corr10}) and
(\ref{eq_imsr1}) unchanged. The dimension $D=8$ operator is 
assumed to be scale invariant. Figure~\ref{fig_running} shows 
the f\/itted observables as a function of $s_0$. The horizontal
bands give the results at $M_\tau^2$ within one standard 
deviation. All curves show a convergent behaviour for 
$s_0\rightarrow M_\tau^2$. Any deviation from the 
f\/itted values for $s_0>M_\tau^2$ should be covered by the ``OPE''
errors assigned to the results~(\ref{eq_l10})--({\ref{eq_o8}).
\vs
Since we use G$\chi$PT formulae in this analysis we have investigated 
the sensitivity of the $(V-A)$ $\tau$ data to a possible constraint 
on the mass ratio $r$ itself. Clearly a combined f\/it of
$L_{10}^{\rm{ef\/f}}$, 
$r$ and the nonperturbative operators must fail due to 
the strong correlations of the input variables which reduce the 
ef\/fective degrees of freedom of the f\/it. Thus, as a test, we may 
use as input for $L_{10}^{\rm{ef\/f}}$ and the nonperturbative operators 
the values~(\ref{eq_l10})--(\ref{eq_o8}) and assume them to be 
perfectly known, \eg, from a precise second measurement. 
F\/ig.~\ref{fig_moments} shows the theoretical 
prediction of the (most sensitive) IMSR moment $R_{\tau,V-A}^{(1,-1)}$ 
as a function of $r$ within the errors from the other theoretical 
sources given in Table~\ref{tab_momdata}, dominated by the error
on $\alpha_s$. Additionally shown as a horizontal band are the ALEPH 
data within experimental errors. We conclude that the current 
experimental precision of the non-strange data does not allow to 
constrain the light quark masses, \ie, the mass ratio $r$. In the 
limit of zero $u,d$ quark masses $(r\rightarrow\infty)$ we obtain 
$R_{\tau,V-A}^{(1,-1)}=5.11$ which is still within the data band of
one experimental and theoretical standard deviation. The sensitivity 
on $r$ when employing the $l\ge0$ moments is even worse than 
with the IMSR.
\begin{figure}[t]
  \epsfxsize12.cm
  \centerline{\epsffile{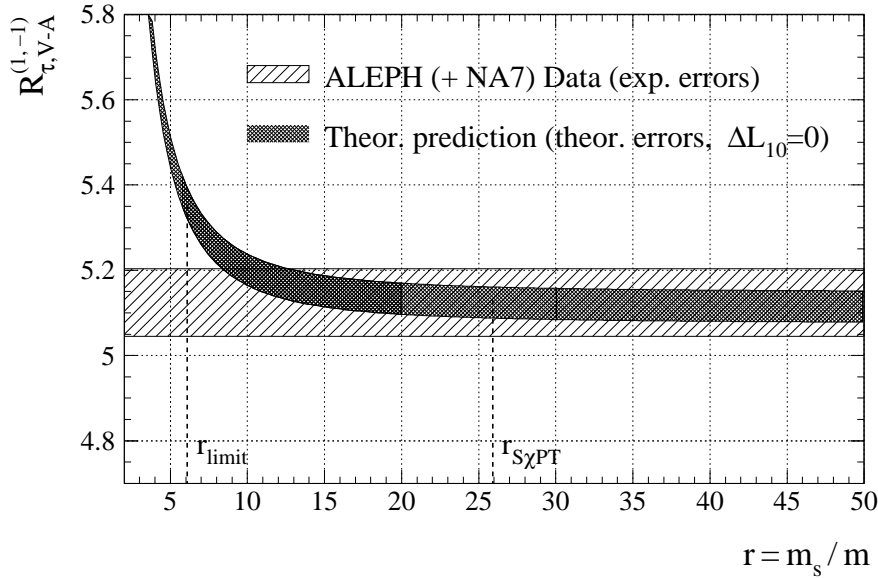}}
  \caption[.]{\label{fig_moments}\it
              Theoretical prediction of the IMSR moment 
              $R_{\tau,V-A}^{(1,-1)}$, using
	      $L_{10}^{\rm{ef\/f}}=-6.36\times10^{-3}$ 
              as fixed input value, versus the 
              mass ratio $r$. The theoretical uncertainty stems mainly 
              from the error on \asm. The dashed band shows the $(V-A)$ 
              data from hadronic $\tau$ decays (including low energy 
              $e^+e^-$ vector cross sections) within experimental errors.}
\end{figure}
\newpage
%
%
\section{Conclusions}

This article deals with a combination of f\/inite energy sum rule 
techniques and Chiral Perturbation Theory ($\chi$PT) low-energy 
expansion in order to exploit recent ALEPH data on the non-strange
$\tau$ vector and axial-vector \sfs\ with respect to an experimental
determination of the $\chi$PT quantity $L_{10}$. The theoretical
predictions of the spectral moments, $R_{\tau,V-A}^{(k,l)}$, of
the $\tau$ hadronic width involve nonperturbative elements of
the Operator Product Expansion when calculating the contour
integral at $|s|=M_\tau^2$. In the case of inverse spectral
moments $(l<0)$, additional $\chi$PT parameters appear
originating from a second contour integral at the $|s|=4M_\pi^2$
production threshold which subtracts the singularity of the
$(s/M_\tau^2)^{-1}$ inverse moment at $s=0$. A constrained f\/it
of $l<0$ and $l\ge0$ spectral moments adjusts simultaneously
the parameter $L_{10}^{\rm{ef\/f}}$, def\/ined by Eq.~(\ref{eq_defl10eff}),
and nonperturbative power operators of dimension $D=6$ and $D=8$. We 
obtain $L_{10}^{\rm{ef\/f}} = -(6.36\pm0.09\pm0.16) \times 10^{-3}$, 
where the f\/irst error is of experimental and the second of theoretical 
origin. The present determination of $L_{10}^{\rm{ef\/f}}$ is 
independent of any chiral expansion; in particular, the value 
obtained here can be directly used in a two-loop analysis: it 
suf\/fices to include higher order corrections in Eq.~(\ref{eq_l10eff}).
Within the one-loop $\chi$PT the above result corresponds to
$L^r_{10}(M_{\rho})=-(5.13\pm0.19)\times10^{-3}$, in good agreement 
with the value  $L^r_{10}(M_{\rho})=-(5.42\pm0.24)\times10^{-3}$ 
extracted from the one-loop analysis of $\pi \rightarrow e\nu\gamma$ 
data and $\langle r^2\rangle_{\pi}$. This provides a non-trivial test
of chiral symmetry underlying $\chi$PT. The total nonperturbative 
prediction to $R_{\tau,V-A}$ found in the f\/it, is in agreement 
with the values of the ALEPH \asm\ analysis~\cite{aleph_asf}. The 
stability of the f\/it results is investigated in performing various
f\/its for ``$\tau$ masses'' smaller than $M_\tau$. Satisfactory 
convergence is observed.
%
%


\newpage
\begin{thebibliography}{99}
\bibitem{wilson}     K.G.~Wilson, \PR\ {\bf 179} (1969) 1499
\bibitem{svz}        M.A.~Shifman, A.L.~Vainshtein and V.I.~Zakharov,
                     \NP\ {\bf B147} (1979) 385, 448, 519
\bibitem{wein}       S.~Weinberg, {\it Physica} {\bf A96} (1979) 327 
\bibitem{gale1}	     J.~Gasser and H.~Leutwyler, {\it Ann. Phys.}  
                     {\bf 158} (1984) 142
\bibitem{chiralsr}   J.F.~Donoghue and E.~Golowich, \PR\ {\bf D49} (1994) 1513
\bibitem{aleph_vsf}  ALEPH \Cl\ (R.~Barate \ea), \ZP\ {\bf C76} (1997) 15
\bibitem{aleph_asf}  ALEPH \Cl\ (R.~Barate \ea), 
                     Report CERN PPE/98-012 (1998), {\it to appear in \EPJC}
\bibitem{dmo}        T.~Das, V.S.~Mathur and S.~Okubo, 
                     \PRL\ {\bf 19} (1967) 895
\bibitem{kagodmo}    E.~Golowich and J.~Kambor, preprint UMHEP-447, ZU-TH
		     30/97, hep-ph/9711256
\bibitem{pichledib}  F.~Le Diberder and A.~Pich, \PL\ {\bf B289} (1992) 165
\bibitem{aleph_as}   D.~Buskulic \ea\ (ALEPH \Cl), \PL\ {\bf B307} (1993) 209
\bibitem{cleo_as}    T.~Coan \ea\ (CLEO \Cl), \PL\ {\bf B356} (1995) 580
\bibitem{imsr}       E.~Golowich and J.~Kambor, \PR\ {\bf D53} (1996) 2651
\bibitem{bnp}        E.~Braaten, S.~Narison and A.~Pich, 
                     \NP\ {\bf B373} (1992) 581
\bibitem{thesis}     A.~H\"ocker, {\it ``Measurement of the $\tau$ \sfs\ and
                     applications to QCD''}, Thesis, Report LAL 97-18, Orsay,
                     France (1997)
\bibitem{chetyrkin}  K.G.~Chetyrkin and A.~Kwiatkowski, \ZP\ {\bf C59}
		     (1993) 525
\bibitem{generalis}  S.C.~Generalis, \JP\ {\bf G15} (1989) L225
\bibitem{chetyrkin2} K.G.~Chetyrkin, S.G.~Gorishny and V.P.~Spiridonov,
                     \PL\ {\bf B160} (1985) 149
\bibitem{lanin}      L.V.~Lanin, V.P.~Spiridonov and K.G.~Chetyrkin,
                     {\it Yad. Fiz.} {\bf 44} (1986) 1372; {\it Sov. J.
                     Nucl. Phys.} {\bf 44} (1986) 892
\bibitem{gale2}      J.~Gasser and H.~Leutwyler, \NP\ {\bf B250} (1985) 465
\bibitem{kago1}      E.~Golowich and J.~Kambor, \NP\ {\bf B447} (1995) 373
\bibitem{kago2}      E.~Golowich and J.~Kambor, preprint ZU-TH-19-97,
 		     hep-ph/9710214
\bibitem{gchpt}      M.~Knecht and J.~Stern, {\it The second
                     DA$\Phi$NE Physics Handbook, Vol~I}, (Eds. L.~Maiani, 
                     G.~Pancheri and N.~Paver), INFN, Frascati (1995) 169
\bibitem{weinberg77} S.~Weinberg, {\it ``A Festschrift for I.I.~Rabi''},
                     Eds. L.~Motz, New York Academy of Sciences, New York (1977)
\bibitem{sterngchpt} N.H.~Fuchs, H.~Sazdjian and J.~Stern, \PL\ {\bf B238}
                     (1990) 380; \PL\ {\bf B269} (1991) 183 
\bibitem{marc}       M.~Knecht, {\it to be published}
\bibitem{sternmainz} J.~Stern, {\it ``Light Quark Masses and Condensates in
		     QCD''}, Invited plenary talk given at the Workshop on
		     Chiral Dynamics: Theory and Experiment (ChPT97), Mainz,
		     Germany, report IPNO/TH-97-30, hep-ph/9712438  
\bibitem{shaomin}    S.~Chen, {\it ``Measurement of $\tau$ decays with kaons 
                     from ALEPH and $m_s$ determination''}, 
                     Talk given at QCD'97, Montpellier, France 1997
\bibitem{jerusalem}  D.~Ward, {\it ``Tests of the Standard Model: W mass 
                     and WWZ Couplings''}, Talk given at the International
                     Europhysics Conference on High-Energy Physics (HEP 97), 
                     Jerusalem, Israel 1997
\bibitem{alphapap}   M.~Davier and A.~H\"ocker, Report LAL 97-85,
                     hep-ph/9711308 (1997), {\it to appear in \PL\ {\it B}}
\bibitem{pdg}        R.M.~Barnett \ea\ (Particle Data Group), 
                     \PR\ {\bf D54} (1996) 1
\bibitem{marciano}   W.J.~Marciano and A.~Sirlin,
                     \PRL\ {\bf 56} (1986) 22; \PRL\ {\bf 61} (1986) 1815
\bibitem{g_2pap}     R.~Alemany, M.~Davier and A.~H\"ocker,
                     Report LAL 97-02 (1997), {\it to appear in \EPJC}
\bibitem{na7}        NA7 \Cl\ (S.R.~Amendolia \ea), \PL\ {\bf B138} (1984) 454
\bibitem{space_like} NA7 \Cl\ (S.R.~Amendolia \ea), \NP\ {\bf B277} (1986) 168
\bibitem{agostini}   G.~D'Agostini, \NIM\ {\bf A346} (1994) 306
\bibitem{georgi}     H.~Georgi, \PL\ {\bf B298} (1993) 187
\bibitem{daphne}     J.~Bijnens, G.~Ecker and J.~Gasser, {\it The second
                     DA$\Phi$NE Physics Handbook, Vol~I}, (Eds. L.~Maiani, 
                     G.~Pancheri and N.~Paver), INFN, Frascati (1995) 125
\bibitem{pienugamma} J.~Bijnens and P.~Talavera, \NP\ {\bf B489} (1997) 387
\bibitem{l1l2}	     L.~Girlanda, M.~Knecht, B.~Moussallam and J.~Stern, \\
		     \PL\ {\bf B409} (1997) 461
\end{thebibliography}
\end{document}